\documentclass[runningheads,citeauthoryear]{apinv}
\usepackage{epsfig,cite,graphics}
\usepackage[T2A]{fontenc}
\usepackage[cp1251]{inputenc}
\usepackage{lscape}
\usepackage{longtable}
\usepackage{hyperref}
\usepackage{url}
\usepackage{amsmath}

\begin{document}

\title{A new prolonged decrease event in the brightness of the young stellar object V2492 Cygni
}
\titlerunning{A new decrease event in the brightness of the YSO V2492 Cyg}
\author{Sunay Ibryamov\inst{1} and Evgeni Semkov\inst{2}}
\authorrunning{S. Ibryamov \& E. Semkov}
\tocauthor{S. Ibryamov \& E. Semkov} 
\institute{Department of Physics and Astronomy, Faculty of Natural Sciences, University of Shumen, 115, Universitetska Str., 9700 Shumen, Bulgaria
	\and Institute of Astronomy and National Astronomical Observatory, Bulgarian Academy of Sciences, 72, Tsarigradsko Shose Blvd., 1784 Sofia, Bulgaria
	\newline
	\email{sibryamov@shu.bg}}
\papertype{Submitted on xx.xx.xxxx; Accepted on xx.xx.xxxx}	
\maketitle

\begin{abstract}
Results from the $BV(RI)_{c}$ photometric observations of the young stellar object V2492 Cyg collected in the period from April 2018 to September 2020 are presented.
These observations are a part of our monitoring of the star that began in 2010 and continuing to date.
V2492 Cyg is located in the Pelican Nebula, and its variability was explained by both accretion and extinction events.
The new photometric data show that the star continues to exhibit rapid irregular variability in all bands.
In the period from March 2019 to May 2020, we registered a prolonged decrease event in the light curve of V2492 Cyg.

\end{abstract}


\section*{1. Introduction}

Studies of the photometric variability of pre-main-sequence (PMS) stars are important to the understanding of the early stages of stellar evolution.
A special kind of variability can be seen in the PMS stars, which undergo outbursts or obscurations.
The outbursts of the PMS stars with amplitude reaching up to 5 mag are grouped into two types: FUors with the prototype FU Orionis (Reipurth \& Aspin 2010) and EXors with the prototype EX Lupi (Herbig 2007).
The outbursts of FUors and EXors are generally attributed to a sizable increase in the accretion rate from the circumstellar disk onto the stellar surface.
The deep minima with amplitude reaching up to 3 mag are mostly seen in the early types of PMS stars, which are called UXors with the prototype UX Orionis (Natta et al. 1997).
It is generally accepted that the observed minima result from the obscuration of the star from circumstellar clouds of dust. 
During the deep minima, the color indices of UXors often become bluer.
This phenomenon is known as the "blueing effect" or "color reverse" (see Bibo \& Th\'{e} 1990).

V2492 Cyg (also know as PTF 10nvg and IRAS 20496+4354) lies in the field of the Pelican Nebula (IC 5070).
The star was discovered by Itagaki \& Yamaoka (2010) during the CCD survey on 23 August 2010 when its unfiltered magnitude was about 13.8.
The authors reported that V2492 Cyg was not detected at its position on a survey reference image taken on 7 August 2008 (limiting magnitude about 17.5).

V2492 Cyg shows the characteristics inherent to EXor- and UXor-type variables.
Detailed studies of the star were made by Covey et al. (2011), K\'{o}sp\'{a}l et al. (2011), Aspin (2011), K\'{o}sp\'{a}l et al. (2013), Hillenbrand et al. (2013), Scholz et al. (2013), Feh\'{e}r et al. (2017), Giannini et al. (2018), Ibryamov et al. (2018).
The star was also subject of several Astronomer's Telegrams (see Semkov \& Peneva 2012, Arkharov et al. 2015, Ibryamov \& Semkov 2017, Munari et al. 2017, Froebrich et al. 2017, Ibryamov \& Semkov 2019).

In our first paper (Ibryamov, Semkov \& Peneva 2018, hereafter ISP18), we presented the results from $BV(RI)_{c}$ observations of V2492 Cyg obtained in the period from 2010 to 2017.
In this study, we present new photometric data of the star and discuss its photometric behavior.

\section*{2. Observations}

The observations of V2492 Cyg were obtained as part of our continuous monitoring of the star (ISP18).
They were collected in the period from April 2018 to September 2020 with the 2-m Ritchey-Chr\'{e}tien-Coud\'{e} (RCC) and the 50/70-cm Schmidt telescopes administered by the Rozhen National Astronomical Observatory in Bulgaria.
Two types of CCD cameras were used to perform of the observations: Andor iKon-L (2048 $\times$ 2048 pixels, 13.5 $\times$ 13.5 $\mu$m pixel$^{-1}$ size and 0.17$\arcsec$ pixel$^{-1}$) on the 2-m RCC telescope, and FLI PL16803 (4096 $\times$ 4096 pixels, 9 $\times$ 9 $\mu$m pixel$^{-1}$ size and 1.08$\arcsec$ pixel$^{-1}$) on the 50/70-cm Schmidt telescope.
The observational procedure and data reduction process was described in ISP18.

All frames were taken through a standard Johnson-Cousins ($BVR_{c}I_{c}$) set of filters.
As a reference, the comparison sequence reported in ISP18 was used.
In very low light, the brightness of V2492 Cyg was under the registration limit of the telescopes used (usually about 20.5 mag for the 2-m RCC telescope, and about 19.5 mag for the 50/70-cm Schmidt telescope).

\section*{3. Results and discussion}

The results of the photometric observations of V2492 Cyg are summarized in Table 1.
The columns contain the Julian date of the observations, the measured magnitudes of the star, and the telescope used.

The $BV(RI)_{c}$ light curves of V2492 Cyg are shown in Fig. 1.
The available data suggest that during our observations, the star shows strong irregular variability in all bands.
It can be seen from Fig. 1 that from April 2018 to January 2019 the star's brightness varies in wide ranges: 12.24-14.00 mag in the $I_{c}$ band, 13.67-15.44 mag in the $R_{c}$ band, 14.94-16.78 mag in the $V$ band, and 16.79-18.88 mag in the $B$ band.
In early March 2019, a deep minimum ($I_{c}$=17.41 mag) in the light curve of V2492 Cyg was observed. 
In April 2019 the brightness of the star began to increase gradually, and at the end of July 2019 it reached $I_{c}$=15.73 mag.

At the beginning of August 2019, the brightness of V2492 Cyg began to decline again, and in mid-January 2020 the deepest drop ($\Delta I_{c}$$=$6.83 mag) was registered.
It is important to mention that during the observations on 15 and 16 January 2020 with the 50/70-cm Schmidt telescope, the star is not visible at its position (Fig. 2).
As can be seen from Fig. 1 at the end of July 2020, V2492 Cyg returned to maximum light.

{\footnotesize
	\begin{longtable}{cccccc|cccccc}
		\caption{Photometric observations of V2492 Cyg in the period April 2018$-$September 2020.}\\
		\hline\hline
		\noalign{\smallskip}  
		J.D.    & $I_{c}$ & $R_{c}$ & $V$   & $B$   & Tel & J.D.    & $I_{c}$ & $R_{c}$ & $V$   & $B$   & Tel\\
		(24...) & [mag]   & [mag]   & [mag] & [mag] &     & (24...) & [mag]   & [mag]   & [mag] & [mag] &    \\
		\noalign{\smallskip}  
		\hline
		\endfirsthead
		\caption{Continued.}\\
		\hline\hline
		\noalign{\smallskip}
		J.D.    & $I_{c}$ & $R_{c}$ & $V$   & $B$   & Tel & J.D.    & $I_{c}$ & $R_{c}$ & $V$   & $B$   & Tel\\
		(24...) & [mag]   & [mag]   & [mag] & [mag] &     & (24...) & [mag]   & [mag]   & [mag] & [mag] &    \\
		\noalign{\smallskip}  
		\hline
		\noalign{\smallskip}  
		\endhead
		\hline
		\label{Tab1}
		\endfoot
		\noalign{\smallskip}		
		58217.573	&	13.13	&	14.72	&	16.07	&	17.97	&	Sch & 58705.349	&	16.90	&	-	    &	-	    &	-	    &	Sch	\\
		58218.560	&	13.11	&	14.71	&	16.04	&	17.96	&	Sch	& 58706.348	&	16.96	&	19.46	&	-	    &	-	    &	Sch	\\
		58220.517	&	13.11	&	14.71	&	16.07	&	18.01	&	Sch	& 58707.438	&	17.06	&	-	    &	-	    &	-	    &	Sch	\\
		58278.385	&	13.41	&	15.03	&	16.41	&	18.31	&	Sch	& 58726.272	&	17.97	&	19.89	&	-	    &	-	    &	2-m	\\
		58281.419	&	13.77	&	15.16	&	16.78	&	18.88	&	2-m	& 58727.333	&	17.92	&	-	    &	-	    &	-	    &	2-m	\\
		58312.385	&	13.46	&	15.28	&	16.76	&	18.66	&	Sch	& 58728.345	&	17.91	&	19.62	&	-	    &	-	    &	2-m	\\
		58316.386	&	12.75	&	14.37	&	15.73	&	17.61	&	Sch	& 58729.296	&	17.96	&	19.74	&	-	    &	-	    &	2-m	\\
		58340.327	&	13.71	&	15.42	&	16.72	&	18.55	&	Sch	& 58729.516	&	18.09	&	-	    &	-	    &	-	    &	Sch	\\
		58342.329	&	13.71	&	15.28	&	16.50	&	18.22	&	Sch	& 58730.301	&	17.90	&	19.57	&	-	    &	-	    &	2-m	\\
		58343.339	&	13.89	&	15.38	&	16.56	&	18.21	&	Sch	& 58730.412	&	17.97	&	-	    &	-	    &	-	    &	Sch	\\
		58344.321	&	13.96	&	15.44	&	16.61	&	18.26	&	Sch	& 58757.344	&	18.73	&	-	    &	-	    &	-	    &	Sch	\\
		58346.273	&	14.00	&	15.27	&	16.50	&	18.15	&	2-m	& 58758.312	&	18.79	&	-	    &	-	    &	-    	&	Sch	\\
		58363.297	&	13.49	&	15.01	&	16.29	&	17.98	&	Sch	& 58759.299	&	18.69	&	-	    &	-	    &	-	    &	Sch	\\
		58364.275	&	13.40	&	14.94	&	16.18	&	17.86	&	Sch	& 58864.233	&	-	    &   -       & -         &   -       &	Sch \\
		58365.498	&	13.31	&	14.63	&	16.04	&	17.79	&	2-m	& 58865.213	&	-	    &   -	    & -         &   -       &	Sch \\
		58409.286	&	12.24	&	13.67	&	14.94	&	16.79	&	Sch	& 58867.200	&	19.07	&	-	    &	-    	&	-	    &	2-m	\\
		58428.234	&	13.35	&	14.88	&	16.17	&	18.00	&	Sch	& 58993.397	&	18.12	&	20.05	&	-	    &	-	    &	2-m	\\
		58435.215	&	12.71	&	14.11	&	15.36	&	17.14	&	Sch	& 58993.421	&	18.44	&	-	    &	-	    &	-	    &	Sch	\\
		58492.176	&	13.42	&	14.77	&	15.97	&	17.67	&	Sch	& 59040.340	&	14.54	&	16.50	&	18.41	&	-	    &	Sch	\\
		58547.594	&	17.41	&	-   	&	-	    &	-	    &	Sch	& 59041.372	&	14.46	&	16.38	&	18.20	&	-	    &	Sch	\\
		58603.449	&	17.13	&	-	    &	-   	&	-    	&	Sch	& 59042.358	&	14.47	&	16.41	&	18.25	&	-	    &	Sch	\\
		58604.474	&	16.96	&	19.29	&	-    	&	-	    &	Sch & 59059.347	&   12.98	&   14.61	&   16.11	&   18.30   &   Sch \\
		58606.408	&	16.72	&	-	    &	-	    &	-	    &	2-m	& 59060.342	&   13.02	&   14.64	&   16.14	&   18.43   &	Sch \\
		58606.483	&	16.76	&	19.29	&	-    	&	-	    &	Sch	& 59075.302	&	13.02	&	14.34	&	15.82	&	17.81   &	2-m \\
		58665.358	&	16.64	&	19.24	&	-	    &	-	    &	Sch	& 59085.259	&	12.78	&	14.20	&	15.50	&	17.45   &	Sch \\
		58666.363	&	16.36	&	18.97	&	-	    &	-	    &	Sch	& 59101.381	&	12.38	&	13.73	&	14.98	&	16.83	&	Sch	\\
		58667.366	&	16.51	&	19.02	&	-	    &	-	    &	Sch	& 59102.320	&	12.40	&	13.57	&	14.75	&	16.70	&	2-m	\\
		58690.282	&	15.82	&	17.59	&	19.66	&	-	    &	2-m	& 59103.306	&	12.58	&	13.78	&	14.99	&	16.92	&	2-m	\\
		58691.386	&	15.98	&	17.74	&	20.02	&	-	    &	2-m	& 59104.355	&	12.31	&	13.64	&	14.85	&	16.67	&	Sch	\\
		58692.337	&	15.73	&	17.46	&	19.83	&	-	    &	2-m	& 59105.283	&	12.36	&	13.68	&	14.91	&	16.74	&	Sch	\\
		58704.317	&	16.91	&	-	    &	-	    &	-	    &	Sch	&           &           &           &           &           &       \\
		\hline \hline
\end{longtable}}

\begin{figure}[]
   \centering
   \includegraphics[width=10cm, angle=0]{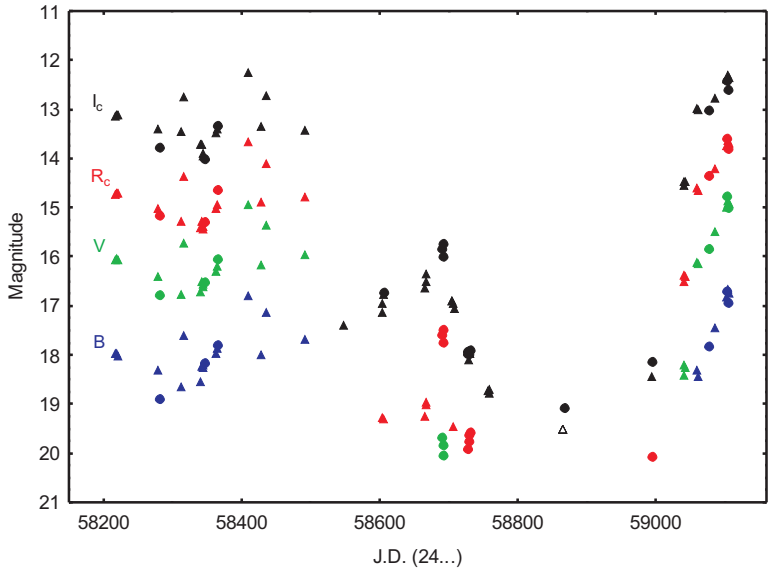}
   \caption{$BV(RI)_{c}$ light curves of V2492 Cyg for the period April 2018$-$September 2020. The circles denote the data from the 2-m RCC telescope, the triangles represent the data from the 50/70-cm Schmidt telescope, and the empty triangles mark the brightness limit in the $I_{c}$ band for two nights (15 and 16 January 2020) where the star was not visible.}\label{Fig1}
   \end{figure}

\begin{figure}[]
   \centering
   \includegraphics[width=4.5cm, angle=0]{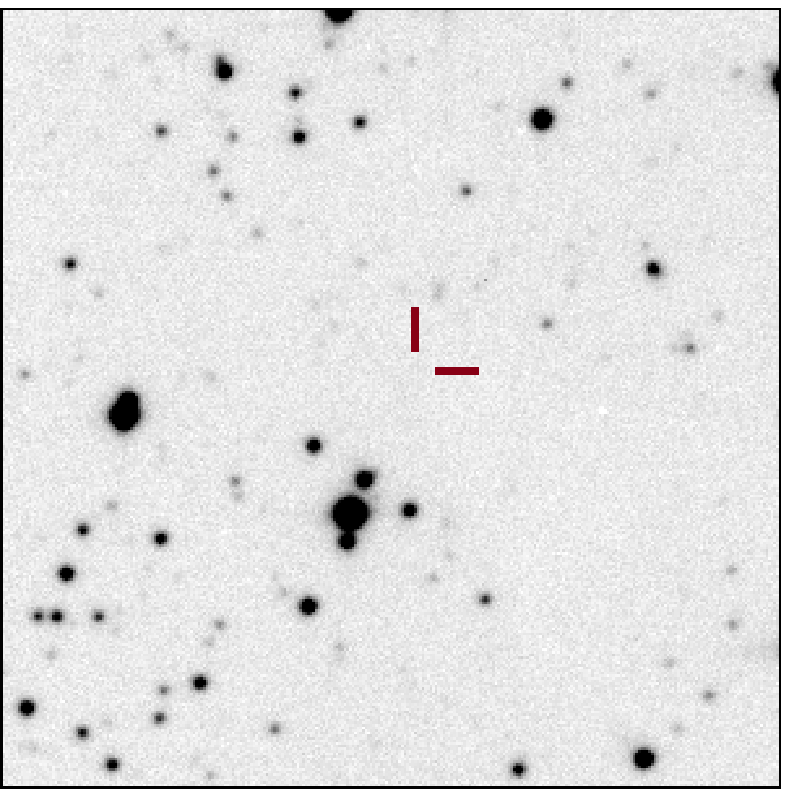}
   \includegraphics[width=4.5cm, angle=0]{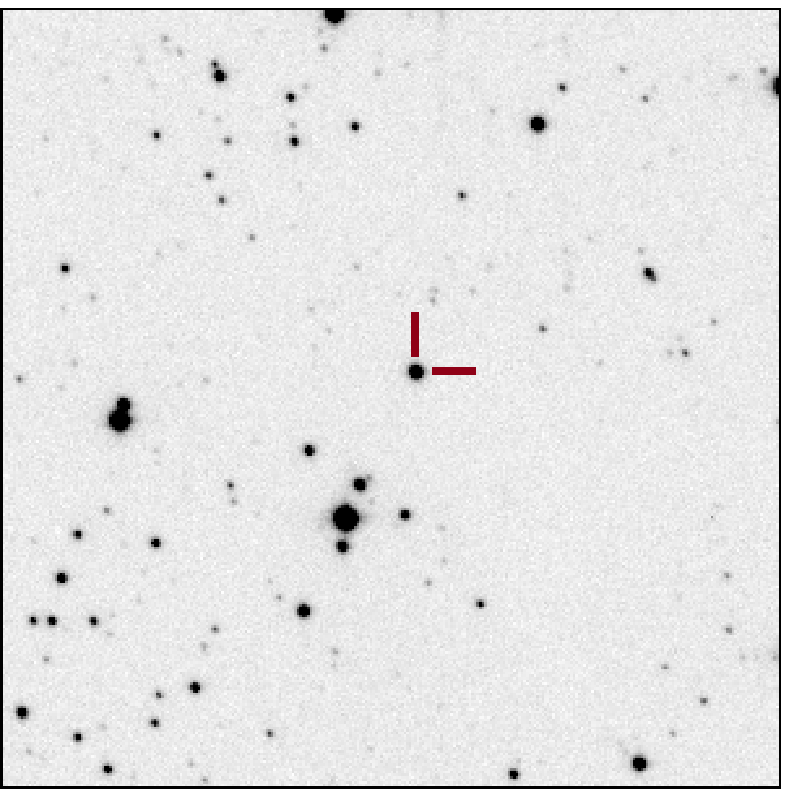}
   \caption{$I_{c}$ frames of the field of V2492 Cyg during the observed deepest drop (16.01.2020, the left image, exp. 300 s) and when the star returned to maximum light (28.07.2020, the right image, exp. 120 s). The position of V2492 Cyg is marked. North is up, east is left. The frames were taken with the 50/70-cm Schmidt telescope. The field of view is 5.5$\arcmin$.}\label{Fig2}
   \end{figure}

We constructed two color$-$magnitude diagrams of V2492 Cyg, which are displayed in Fig. 3.
The diagrams indicate that the star becomes redder when fainter in a manner consistent with changing extinction.
Such color variations are typical for PMS stars, whose light is covered by dust grains or filaments in the line of sight.
In the color indices, there is no blue turning during the low states of the star.
In this case, it can be assumed that extinction is not the predominant mechanism for the brightness variation of V2492 Cyg.

\begin{figure}[]
	\begin{center}
		\includegraphics[width=7cm]{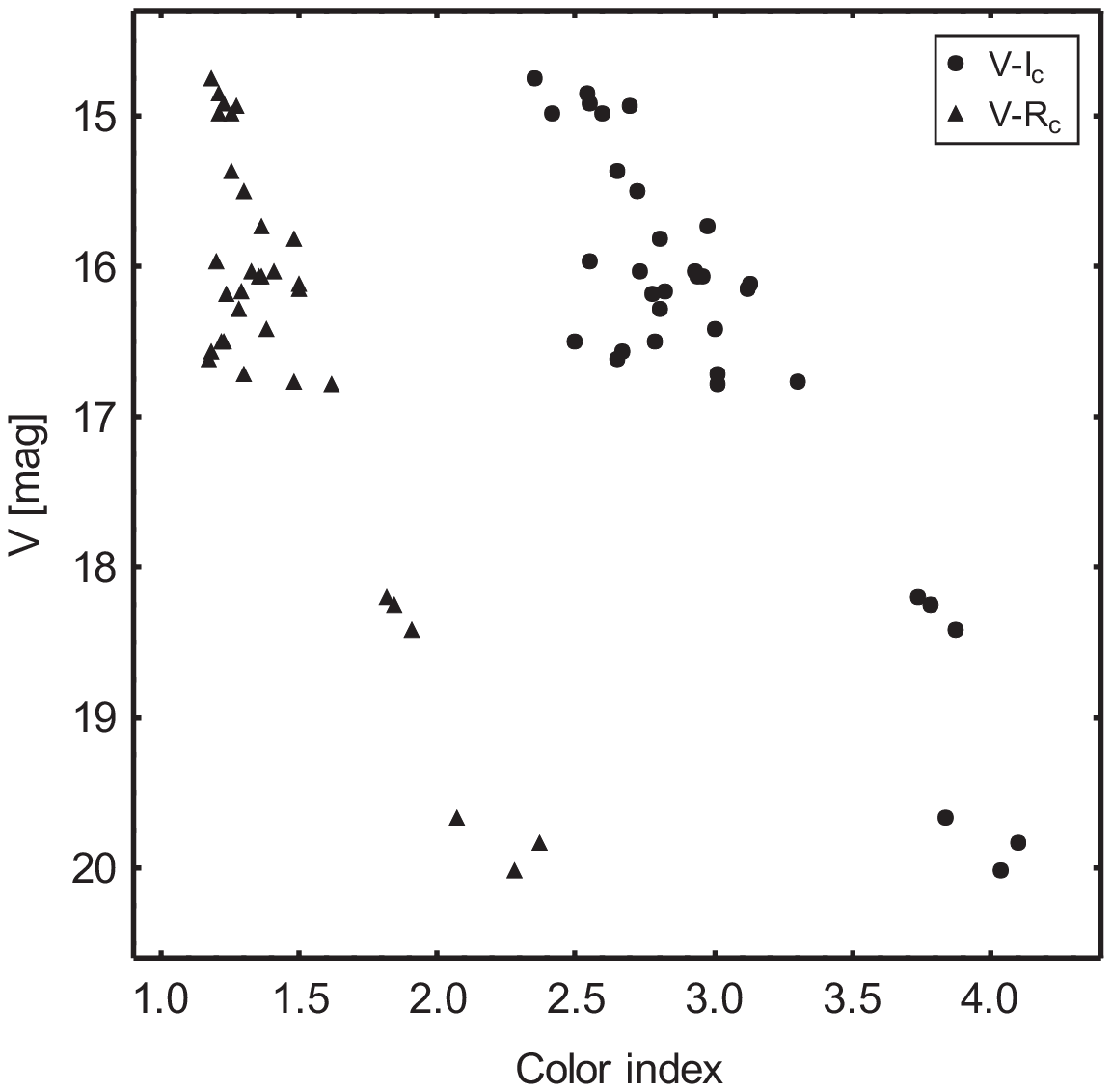}
		\caption{Color indices $V-I_{c}$ and $V-R_{c}$ versus the stellar $V$ magnitude of V2492 Cyg for the period April 2018$-$September 2020.}\label{Fig3}
	\end{center}
\end{figure}

The long-term $I_{c}$ light curve of V2492 Cyg from all available observations is shown in Fig. 4.
It can be seen that the star shows large-amplitude variations in the brightness.
Several clearly expressed deep minima with different amplitudes and duration in the light curve can be distinguished.
The variability of the star may result from the superposition of both variable accretion rate and variable extinction, although their mutual proportion has not been clarified yet.
It is very difficult to distinguish the two phenomena using only photometric data.
We carried out a periodicity search of the deep minima in the brightness of V2492 Cyg, but we did not find any a reliable one.

\begin{figure}[]
   \centering
   \includegraphics[width=\textwidth, angle=0]{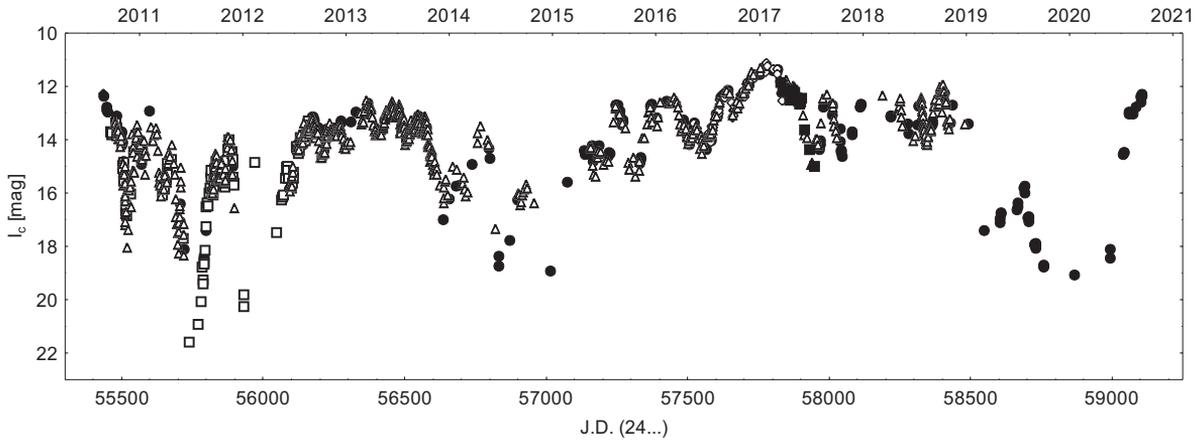}
   \caption{Long-term $I_{c}$ light curve of V2492 Cyg for the period August 2010$-$September 2020. The meanings of different symbols used are as in ISP18 (Fig. 2 therein).}\label{Fig4}
   \end{figure}

\section*{ 4. Conclusion}

We presented and discussed the results from photometric observations of V2492 Cyg during the observed new decrease event in its brightness.
The star shows strong irregular variability in all bands.
This variability can be likely explained by a superposition of both variable accretion rate and variable extinction, although their mutual proportion has not been clarified yet.
It is difficult to distinguish the two phenomena using only photometric data.
V2492 Cyg retains its complicated light curve, and further observations (photometric and spectroscopic) are very important for clarifying its exact nature.

\section*{Acknowledgements}

This work was supported by the Bulgarian Ministry of Education and Science under the National Program for Research "Young Scientists and Postdoctoral Students".
This research has made use of NASA's Astrophysics Data System Bibliographic Services.

\end{document}